%% file: Fixed_point_hierarchical_arxiv.tex
\documentclass[review,dvipsnames]{elsart}
\usepackage{bm,xcolor,relsize}
\usepackage{tikz}

\usepackage[utf8]{inputenc}

\usetikzlibrary{arrows,shapes}
\usetikzlibrary{mindmap,trees}
\usepackage{pgfplots}
\usetikzlibrary{positioning}
\usepackage{ifthen}
\usetikzlibrary{patterns}
\usetikzlibrary{calc}

\usepackage{graphicx,amsmath,amssymb,here}
\journal{Journal of Process Control}
\begin{document}
\begin{frontmatter}
\title{Fixed-point Based Hierarchical MPC Control Design For a Cryogenic Refrigerator}
\author[gipsa]{M. Alamir}\ead{mazen.alamir@gipsa-lab.fr}
\author[sbt]{P. Bonnay}\ead{patrick.bonnay@cea.fr}
\author[sbt]{F. Bonne}\ead{francois.bonne@cea.fr}
\author[gipsa]{V. V. Trinh}\ead{vanvuong.trinh@gipsa-lab.fr}
\address[gipsa]{CNRS, Univ. Grenoble Alpes, Gipsa-lab, F-3800, Grenoble, France.}
\address[sbt]{Univ. Grenoble Alpes, INAC-SBT, F-38000 Grenoble, France.}
\thanks{This work has been funded by the French ANR (Agence Nationale pour la recherche) project CRYOGREEN.}
\begin{abstract}
In this paper, a simple and general hierarchical control framework is proposed and validated through the interconnection of the Joule-Thomson and the Brayton cycle stages of a cryogenic refrigerator. The proposed framework enables to handle the case of destabilizing interconnections through state and/or control signals (which is the case of the cryogenic refrigerator example). Moreover, it offers the possibility to simply change the behavior of the overall system (depending on the context) by only changing the coordinator problem's parameters without changing the set of local controllers used by subsystems which is a common industrial requirement regarding industrial control architectures. Finally, the proposed scheme enables a smooth operator handover on a specific subsystem and/or actuator. 
\end{abstract}
\end{frontmatter}
\maketitle
\section{Introduction} \label{secintro}
Nowadays it is widely admitted that non centralized control architectures are to be preferred, when possible, to fully centralized ones \cite{Scattolini2009,Negenborn2014}. Many reasons for this can be invoked including breaking large problems into small tractable ones, robustness, modularity, privacy preservation and easiness of maintenance. 

Non centralized architectures can be mainly divided into distributed and hierarchical. In the former, no coordinator is used and communication takes place  between adjacent neighbors while in the latter, subsystems communicate exclusively with a coordinator which attempts to coordinate the subsystems behavior in order to achieve a global task or to optimize some globally defined performance index. 

Excellent reviews of the state of the art regarding these architectures are given in \cite{Scattolini2009,Negenborn2014} as far as Model Predictive Control (MPC) design is concerned. These reviews show a highly dynamic research areas. In particular, the survey book \cite{Negenborn2014} enumerates no less than 35 distinguishable approaches. This makes hard for any new proposed solution to claim full novelty. Instead, let us summarize the items that might help positioning the hierarchical solution proposed in this paper w.r.t existing works on hierarchical control architectures\footnote{As far as process control is concerned, hierarchical architectures are preferred as there is always some global goal to achieve such as global efficiency and/or product quality \cite{Qin2003,Seborg2004}. Distributed architectures are more adapted to manage the behavior of a set of similar agents (robots, vehicles, drones, etc.) although some distributed schemes can also be used in process control.}:
\begin{itemize}
\item[$\checkmark$] Many works focus on enforcing the optimality  and/or constraints fulfillment by assuming the existence of a fully decentralized {\bf stabilizing} feedback \cite{Barcelli2010,DoaKev:13-001,Doan2011}. This strong assumption is not required here nor is it valid for the cryogenic process under interest. On the other hand, our stability result does not theoretically hold in the presence of constraints although it can be technically applied with potential success. This obviously makes the two families of solutions {\em non comparable} as on one hand, many processes does not fit the fully centralized stabilizability assumption  while on the other hand, many processes need constraints to be explicitly handled. Now it is a fact that the situation in the process industry leaves dozens if not hundreds of non coordinated PID-like controllers that do not address the constraints anyway nor they guarantee any kind of stability even in the presence of relevant scenarios. \\
\item[$\checkmark$] In some works, a dedicated assumptions regarding the nature of the coupling signals are introduced in order to avoid strong destabilizing coupling effects. For instance, the framework of \cite{Stewart2010, Ocampo2012} assumes only coupling through control input actions while in \cite{Picasso2010} a two layer hierarchical structure is studied where the the higher layer system is slow with control input given by the lower level. Such assumptions circumvent the major difficulties in strongly coupled systems where a typical {\bf fixed-point} iterations lies beneath with a potential risk of non convergence during the iterations between the coordinator and the subsystems. This mechanism is clearly highlighted in the paper and its convergence is analyzed. \\
\item[$\checkmark$] A large part of the literature borrows the viewpoint according to which non centralized frameworks are mainly useful to break down the complexity/dimensionality of the centralized problem. In this perspective, the full knowledge of the system is still assumed in the coordinator level but the computation is broken into several parallel parts that are achieved by the subsystems. This approach does not address an important industrial concern according to which, it should be possible to change one subsystem locally (switch to a new generation of valves or change the logic of the local controller from PID to MPC for instance) without being obliged to propagate the consequences of this change to every line of algorithm at the remaining local controllers and/or the coordinator's level. \\
\item[$\checkmark$] Finally, putting aside the always questionable novelty assertion with regards to the huge literature on the topics, it is a fact that too many proposed schemes are validated through toy examples such as steered tank reactor or coupled water tanks, etc\footnote{A nice counter example, among a few  others, where a relevant and challenging example is handled can be found in \cite{Ocampo2012} where hierarchical framework is applied to the Barcelona's drinking water network.}. The example addressed in this contribution is quite significant: two coupled systems with $10$ and $14$ states respectively, each one acts on the other through $3$ coupling signals including destabilizing coupling effect. All these characteristics make the example a nice benchmark for existing and future hierarchical control architectures and algorithms. To this respect, the paper can be viewed as an introduction to such a benchmark together with a first successful hierarchical control architecture that might be compared to future candidate hierarchical design.
\end{itemize} 
This paper is organized as follows: First of all, the system under study is presented and the control objective is explained in Section \ref{seccryo}. This presentation enables the general setting and the proposed solution sketched in Section \ref{secgeneral} to be better understood. Section \ref{secresult} gives the results obtained by means of the proposed hierarchical framework on the cryogenic refrigerator. Finally, Section \ref{secdiscussion} gives further discussion regarding the proposed solution and how it can be used to smoothly handle the case of operator handover at several levels. Finally, the paper ends by Section \ref{secconclusion} which summarizes the paper and gives some hints for further investigations. 
\section{Problem description} \label{seccryo} 
\subsection{Presentation of the cryogenic refrigerator}
\begin{figure}
\begin{center}
\includegraphics[width=\textwidth]{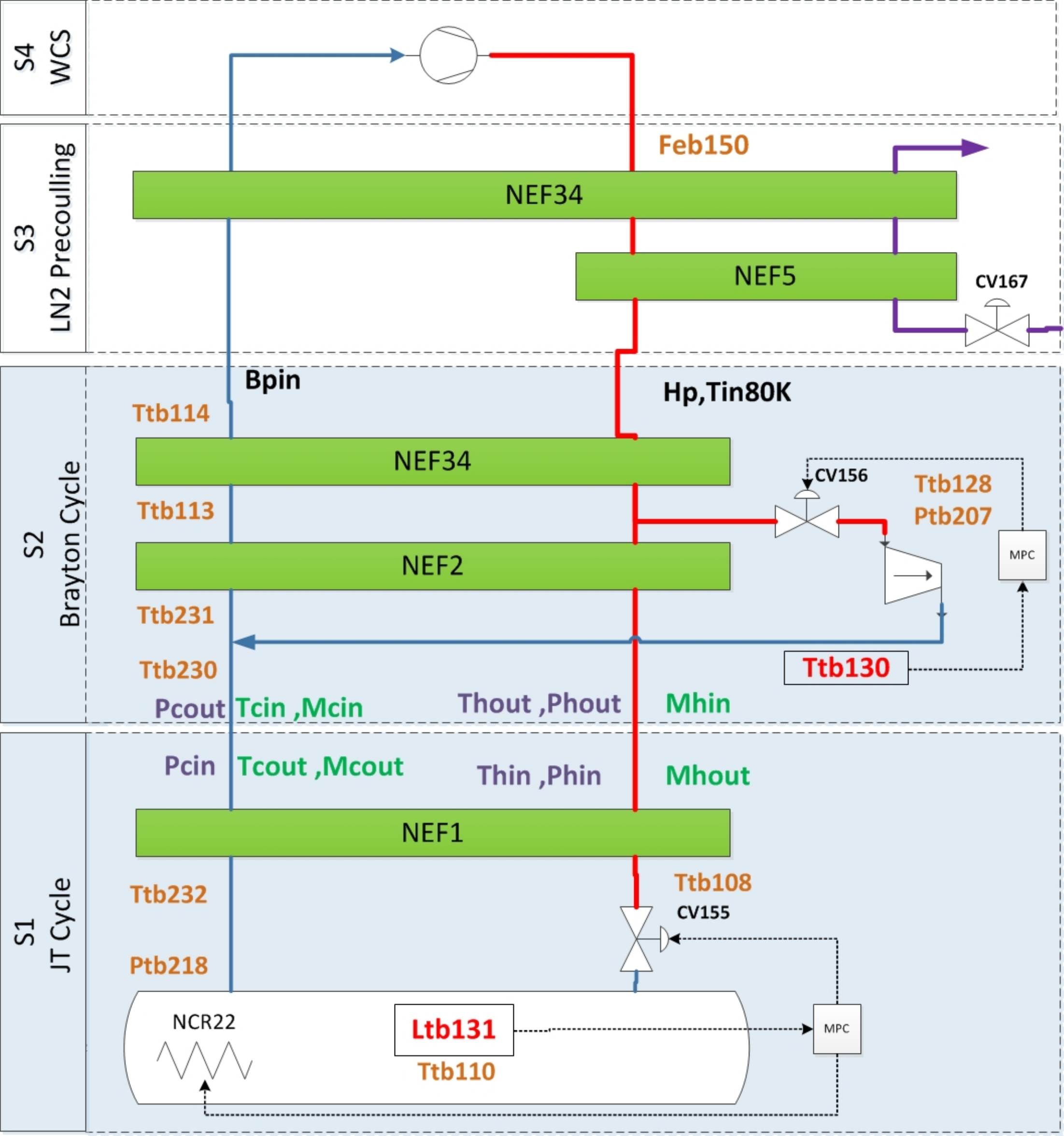} 
\end{center} 
\caption{Sketch of the cryogenic refrigerator of CEA-SBT including the two coupled subsystems: $S_1$: The Joule-Thomson cycle. $S_2$: The Brayton cycle. Figure \ref{SBTstation} shows views of the real system. } \label{thesystem} 
\end{figure}

Cryogenic refrigerators are used in the experimental facilities containing supra-conducting circuits in order to provide cooling power to cool them down. Figure \ref{thesystem} shows a sketch of the 400W@1.8 K experimental refrigerator (in the 400W@4.4 K configuration) at CEA\footnote{Commissariat à l'Energie Atomique} /INAC\footnote{Institut Nanoscience et Cryogénie}/SBT\footnote{Service des Basses Températures} \cite{Roussel2006}. This figure shows the system to be controlled which is composed of subsystems $S_1$ (The Joule-Thomson cycle) and $S_2$ (The Brayton cycle). 
Generally speaking, a cryogenic refrigerator implements a thermodynamic cycle which produces cooling power by extracting work from a fluid by means of cryogenic turbine (the one following the valve CV$_{156}$ in $S_2$ of Figure \ref{thesystem}) and by exchanging heat power through a series of heat exchangers (denoted by NEF$_x$ in Figure \ref{thesystem}). The main objective is to absorb the disturbing heat power generated by the operation of the experimental facility which is represented in the 400W experimental facility by the heat source denoted by NCR$_{22}$ in system $S_1$ of Figure \ref{thesystem}. 

Note that in order to close the thermodynamic cycle, a compressor is used in the so-called warm zone (composed of subsystems $S_3$ and $S_4$ on Figure \ref{thesystem}) with the objective to maintain pressures (denoted by $H_p$, $B_p$ in Figure \ref{thesystem}) at some prescribed level which is presumed to adapted to a well-conditioning of the overall system. For more details regarding the principle and the modeling of cryogenic plants, interested reader can refer to \cite{Bradu2009,Bonne2015}.

In this contribution, it is assumed that the warm zone (systems $S_3$ and $S_4$) is appropriately controlled (see for instance \cite{Bonne2013}) and we focus on the control of the so-called cold zone including subsystems $S_1$ and $S_2$. The corresponding control problem is described in the next section.
\subsection{The control problem of the cold zone $S_1$-$S_2$}
Using the notation of Figure \ref{thesystem}, the control problem of the cold zone including $S_1$ and $S_2$ is described hereafter by successively describing the control inputs, the regulated outputs, the dynamical model and the control objective:
\begin{figure}
\begin{center}
\includegraphics[width=\textwidth]{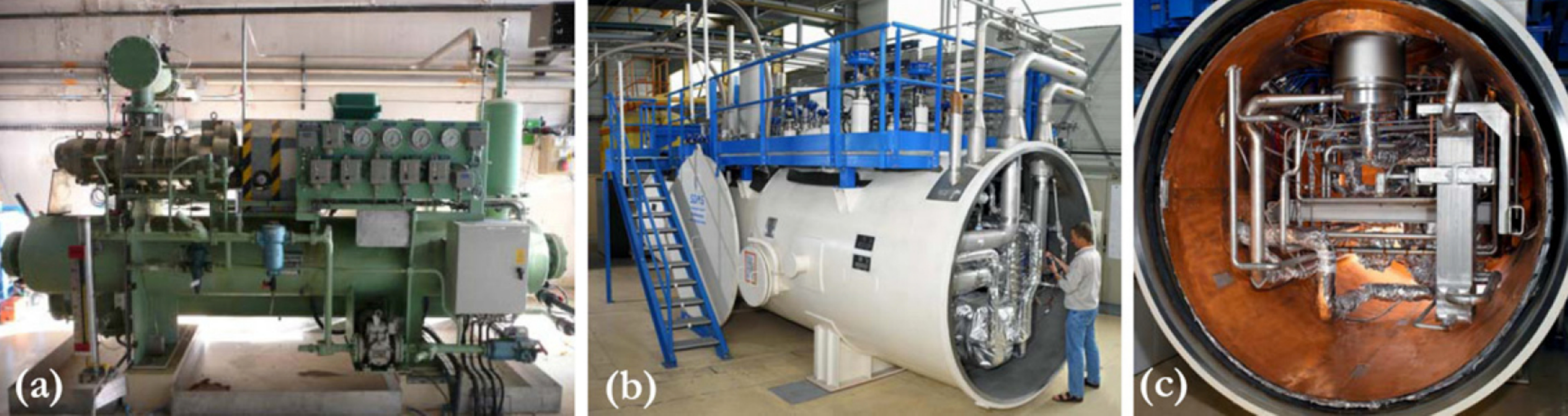} 
\end{center} 
\caption{View of the cryogenic plant of CEA-INAC-SBT, Grenoble. (a) The compressor of the warm compression zone (b) Global view of the cold zone (c) details of the cold zone.} \label{SBTstation} 
\end{figure}

\subsubsection{The control inputs}
There are three control inputs, two of them belong to the Joule-Thomson cycle ($S_1$) and one belongs to the Brayton cycle ($S_2$), more precisely (see Figure \ref{thesystem}):
\begin{enumerate}
\item CV$_{155}$: the position ($\in [0,100]$) of the valve situated at the input of the Helium bath (Figure \ref{thesystem}). This valve is referred to as the JT-valve.\\
\item NCR$_{22}^{(1)}$ : the heating power (W) of the resistance inside the Helium bath. Recall that this same actuator is used to simulate heat pulses coming from the operation of the physical experimental facility served by the cryogenic refrigerator. This explains the upper index used in NCR$^{(1)}_{22}$ as the whole power is split into two components NCR$_{22}$=NCR$^{(a)}_{22}$+NCR$^{(w)}_{22}$, the first of which is used as a control input inside $S_1$ while the second is used as a disturbance signal. Note that although an observer can be built to estimate the disturbance NCR$^{(w)}_{22}$ (see \cite{Bonne2014JPC} for instance), this disturbance is considered here as an unmeasured disturbance because of the difficulties to predict its future behavior in an MPC design\footnote{Indeed, the temporal characteristics of the heat pulse depend on the experiment being conducted on the experimental facility. This is required to be unknown by the cryogenic refrigerator.}.\\
\item CV$_{156}$: the position ($\in [0,100]$) of the valve situated at the inlet of the cryogenic turbine of $S_2$. 
\end{enumerate} 
\subsubsection{The regulated outputs}
Three output variables need to be regulated, as many as control inputs in each subsystem. More precisely:
\begin{enumerate}
\item Ltb$_{131}$: the level (\%) in the liquid helium bath. This variable represents the thermal storage that is immediately available for cooling purposes.\\
\item Ttb$_{108}$: the temperature ($K$) at the inlet of the JT-valve in subsystem $S_1$ (see Figure \ref{thesystem}). The tight regulation of this variable is crucial in order to maintain the efficiency of the process. \\
\item Ttb$_{130}$: the temperature ($K$) at the outlet of the cryogenic turbine. Maintaining this temperature above a lower limit is mandatory in order to avoid the formation of liquid droplets that can break the turbine. 
\end{enumerate} 
\subsection{The nominal operation point}
A nominal operation point is defined which is considered to be optimal given the sizing of the refrigerator. This operational point is defined by the stationary values of the inputs, outputs and disturbance described above, namely:\\
\begin{equation}
\mbox{\rm CV$_{155}^0$, NCR$_{22}^{(a,0)}$, CV$_{156}^0$, Ltb$_{131}^0$, Ttb$_{108}^0$, Ttb$_{130}^0$ and NCR$_{22}^{(w,0)}$}
\end{equation} 
These values enable the deviations in the inputs, outputs and disturbances to be defined according to:
\begin{equation}
u_1:= \begin{bmatrix}
CV_{155}-CV_{155}^0\cr 
NCR_{22}^{(a)}-NCR_{22}^{(a,0)}
\end{bmatrix} \quad;\quad u_2:=CV_{156}-CV_{156}^0
\end{equation} 
\begin{equation}
y_1:= \begin{bmatrix}
Ltb_{131}-Ltb_{131}^0\cr 
Ttb_{108}-Ttb_{108}^0
\end{bmatrix}  x\quad;\quad y_2:=Ttb_{130}-Ttb_{130}^0
\end{equation} 
and 
\begin{equation}
w=NCR_{22}^{(w)}-NCR_{22}^{(w,0)}
\end{equation} 
In what follows, the following notation is used to denote the values at the operation point:
\begin{equation}
U_1^0:= \begin{bmatrix}
CV_{155}^0\cr NCR_{22}^{(a,0)}
\end{bmatrix}  \ ,\ U_2^0:=CV_{156}^0\ ,\ Y_1^0:= \begin{bmatrix}
Ltb_{131}^0\cr Ttb_{108}^0
\end{bmatrix}  \ ,\ Y_2^0:=Ttb_{130}^0
\end{equation} 
\subsubsection{The dynamic model}
Using the index $s\in \{1,2\}$ to refer to the subsystem $S_s$, the linearized model of the two subsystems, defined around the operation point described above takes the form:
\begin{align}
x_1^+&=A_1x_1+B_1u_1+G_1v_1+F_1w_1\label{systx1}\\ 
x_2^+&=A_2x_2+B_2u_2+G_2v_2+F_2w_2\label{systx2}\\
v_1&=C_{v_1}x_2+D_{v_1}u_2+E_{v_1}v_2 \label{systv1} \\
v_2&=C_{v_2}x_1+D_{v_2}u_1+E_{v_2}v_1 \label{systv2}\\
y_1&=C_1x_1+D_1u_1+E_1v_1 \label{systy1} \\
y_2&=C_2x_2+D_2u_2+E_2v_2 \label{systy2} 
\end{align} 
in which 
\begin{itemize}
\item[$\checkmark$] $x_1\in \mathbb{R}^{10}$ and $x_2\in \mathbb{R}^{14}$ are the states of subsystems $S_1$ and $S_2$ respectively.
\item[$\checkmark$] $y_1\in \mathbb{R}^{2}$ and $y_2\in \mathbb{R}$ are the deviations of regulated outputs described above. 
\item[$\checkmark$] $v_1\in \mathbb{R}^{3}$ and $v_2\in \mathbb{R}^{3}$ are coupling signals that make the two systems interact dynamically as $v_1$ depends on $(x_2,u_2)$ while $v_2$ depends on $(x_1,u_1)$. As a matter of fact $v_1$ and $v_2$ both depend on the whole state $x$ and the control $u$ as $v_1$ depends also on $v_2$ and vice-versa [see (\ref{systv1}) and (\ref{systv2})]. 
\end{itemize} More precisely, using the notation of Figure \ref{thesystem}, the coupling variable $v_1$ is given by
\begin{equation}
v_1:=(P_h,T_{h},P_c)^T\in \mathbb{R}^{3}
\end{equation} 
where $P_h$ and $T_{h}$ are the pressure and temperature (deviations) at the downstream inlet of the heat exchanger NEF1 while $P_c$ stands for the pressure (deviation) at the upstream outlet of the same heat exchanger. 

Similarly, the coupling variable $v_2$ is given by:
\begin{equation}
v_2:=(M_h,T_{c},M_c)^T\in \mathbb{R}^{3}
\end{equation} 
where $M_h$ is the mass flow rate (deviation) at the downstream inlet of the heat exchanger NEF1 while $T_c$ and $M_c$ stand respectively for the temperature and the mass flow rate (deviations) at the upstream outlet of the same heat exchanger.  

From these equations, it comes clearly that the two subsystems $S_1$ and $S_2$ are strongly coupled since the dynamics of $S_1$ defined by (\ref{systx1}) depends on the variable $v_1$ which depends through (\ref{systv1}) on $x_2$, $u_2$. The same can be said about the dynamics of $S_2$ which depends on $S_1$ through the coupling variable $v_2$ which depends through (\ref{systv2}) on $x_1$ and $u_1$.

In the sequel, the equations (\ref{systx1})-(\ref{systy2}) are sometimes shortly rewritten in the following compact form when linearity is not explicitly involved:
\begin{align}
x_s^+&=f_s(x_s,u_s,v_s,w_s) & s=1,2 \label{shortxs}\\
v_1&=g_1(x_2,u_2,v_2) \label{shortv1} \\
v_2&=g_2(x_1,u_1,v_1) \label{shortv2} \\
y_s&=h_s(x_s,u_s,v_s,w_s) & s=1,2 \label{shortys}
\end{align} 
Note that for any initial state $x(k)$ and any control profile $\bm u$ defined over some prediction horizon of length $N$, the corresponding nominal [disturbance-free] trajectories $\bm X(\cdot,\bm u,x(k)\vert \bm v)$ that are obtained by integrating the dynamics (\ref{shortxs})-(\ref{shortv2}) enables the following so-called coherence constraints to be defined:
\begin{align*}
\bm v_1(k+i)&=g_1\Bigl(\bm X_2(k+i,\bm u_2,x_2(k)\vert \bm v_2),\bm u_2(k+i),\bm v_2(k+i)\Bigr) \quad \forall i\in \{1,\dots,N\}\\
\bm v_2(k+i)&=g_2\Bigl(\bm X_1(k+i,\bm u_1,x_1(k)\vert \bm v_1),\bm u_1(k+i),\bm v_1(k+i)\Bigr) \quad \forall i\in \{1,\dots,N\}
\end{align*} 
which are obviously two conditions on the coupling variable profiles $\bm v:=(\bm v_1,\bm v_2)$ to be compatible with the system's nominal (disturbance-free) coupled equations. This condition can be shortly written as follows:
\begin{align}
\bm v_1=\bm g_1(\bm u_2,x_2(k),\bm v_2) \label{coherenceeq1}\\ 
\bm v_2=\bm g_2(\bm u_1,x_1(k),\bm v_1) \label{coherenceeq2}
\end{align} 
with obvious appropriate notation. These last conditions are referred to as the {\bf coherence constraints}.
IN what follows, the obvious notation $x=(x_1^T,x_2^T)^T$, $y=(y_1^T,y_2^T)^T$, $v=(v_1^T,v_2^T)^T$  are used. 
\subsubsection{The control objective}
There are two modes that define the control objective:
\begin{enumerate}
\item In the first mode which is the by-default mode, the control has to regulate the system $S_1\cup S_2$ around $x=0$ (the nominal steady state operation point invoked above) despite of the presence of non measured disturbance. This is a disturbance rejection mode representing the {\em raison d'être}  of the cryogenic refrigerator. \\ 
\item In the second mode, the operator should be able to temporary steer the system to a different steady state that corresponds to some new set-point $y\neq 0$. For instance, the operator would like to change the level of the liquid helium in the bath (new value of the set-point on Ltb$_{131}$) or on the critical temperature Ttb$_{108}$. 
\end{enumerate} 
These two modes can be accounted for by using different set-points and different weighting matrices in the following {\em centralized}  cost function: 
\begin{equation}
J_c(\bm u,x(k)):=\sum_{s=1}^2\left[\sum_{i=1}^N\left(\dfrac{i}{N}\right)^q\left[\left\|y_s(k+i)-r_s^d\right\|_{Q_c^{(s)}}^2+\left\|U_s^0+u_s(k+i)\right\|_{R_c^{(s)}}^2\right]\right] \label{centralcost} 
\end{equation} 
where $\bm u$ is a candidate sequence of control inputs to be applied starting from the initial state $x(k)$ over the prediction horizon $[k,k+N]$. Note that the output set-points $r_s^d$, for $s=1,2$ represent the desired values of the deviation on the output corresponding to the set-point $r_s^d+Y_s^0$. Finally, $Q_c^{(1)}\in \mathbb{R}^{2\times 2}$, $Q_c^{(2)}\in \mathbb{R}_+$, $R_c^{(1)}\in \mathbb{R}^{2\times 2}$ and  $R_c^{(2)}\in \mathbb{R}^{+}$ are non negative weighting matrices. 
\begin{rem}
Note that the time-dependent weighting term $(i/N)^q$ for some $q\in \mathbb N$ enables to put higher weight on the tail of the prediction horizon by taking high value of $q$. Note that taking $q=0$ recover the standard time-invariant stage cost. 
\end{rem}
\begin{rem}
It is worth underlining that in the cost function (\ref{centralcost}), the total value of the control inputs are penalized and not the deviation between the input and the steady input that is typically used in a standard Model Predictive Control formulation. This is because in common MPC formulations where the control is penalized through a term of the form $\|u_s(k+i)-u_s^d\|$, the stage cost at the desired position is always $0$ which obviously does not represent the difference in the real cost that might correspond to each different set-point. Another difference lies in the fact that in order to guarantee the stability of MPC scheme, the whole state should be penalized using a term of the form $\|x_s(k+i)-x_s^d(r_s^d))\|$ (see \cite{Mayne2000} for more details). This again leads to a cost function that does not rigorously represent the economic/performance criterion one would intuitively like to minimize. We will see that such formulations which are necessary for the stability of the resulting MPC will be used in the formulation of the local MPC controllers while the realistic and relevant cost function (\ref{centralcost}) is used to formulate the centralized problem's cost. Such considerations are intimately linked to the concept of economic Model Predictive Control \cite{Rawlings2012} although the latter is generally studied in a non hierarchical framework. 
\end{rem}
Based on the above definitions and notation, the control problem can be stated as follows:

\begin{minipage}{0.2\textwidth}
\bf Problem  statement
\end{minipage} 
\begin{minipage}{0.02\textwidth}
\rule{0.5mm}{85mm}
\end{minipage} 
\begin{minipage}{0.8\textwidth}
Define a hierarchical control scheme in which two local MPC controllers, defined for $S_1$ and $S_2$ respectively, receive appropriate set-points $r_1$ and $r_2$ (send by a coordinator) that minimize the centralized cost function (\ref{centralcost}) defined for some desired set-point $r^d$. Moreover, the following conditions should be satisfied:\\
\begin{enumerate}
\item $S_1$ and $S_2$ exclusively communicate with the coordinator \\
\item The coordinator ignores the details of the mathematical models inside $S_1$ and $S_2$ and the details of their controllers\\
\item The different modes described above are handled by simply changing the desired set-point $r^d$ and/or the weighting matrices $Q_c^{(s)}$ and $R_c^{(s)}$ of the centralized problem without changing the tuning of the local controllers. 
\end{enumerate} 
\end{minipage} 

As it is explained later on, there is generally no reason to have $r=r^d$ as the local costs and the central sub-costs are quite different. The optimal local set-points $r_s$, $s=1,2$ are to be computed by the coordinator in order to minimize the central cost corresponding to the original set-point $r^d$.

In the next section, a solution is proposed for the above hierarchical control problem in the general case before it is validated on the cryogenic refrigerator in Section \ref{secresult}.  It is worth underlying that while the presentation of the next section considers only two subsystems for the sake of clarity and in order to simplify the notation, the framework is obviously applicable almost without change to the case of a networked of several coupled subsystems. Only the notation would be heavier which we prefer to avoid here. 
\section{General setting and proposed solution} \label{secgeneral} 
In this section, it is first shown (section \ref{subsecestim}) that a fixed point-like iteration can be defined so that if convergence arises, the coordinator can get a hierarchical estimation of the central cost for a candidate value $r=(r_1,r_2)$ of the set-point to be sent to the subsystems. For the sake of clarity of exposition, it is first assumed that this convergence occurs systematically and the way the loop is closed to get a hierarchical feedback control addressing the hierarchical control problem stated in the previous section is shown (Section \ref{subsectionclosetheloop}). The section is then ended by the analysis of the convergence of the fixed point iteration (Section \ref{subsectionCA}).
\subsection{Hierarchical estimation of the central cost by fixed point negotiation} \label{subsecestim} 
Let us consider two subsystems $S_1$ and $S_2$ which are described by the coupled system of equations given by (\ref{shortxs})-(\ref{shortv2}). Consider also the hierarchical control problem stated at the end of the previous section with the central cost (\ref{centralcost}). During this subsection, the states $x_s(k)$, the central set-point $r_s^d$ and the auxiliary individual set-points $r_s$ are supposed to be given and frozen.

Note first of all that the central optimization problem (\ref{centralcost}) can be redefined by considering the extended vector of degrees of freedom $(\bm u,\bm v)$ as follows:
\begin{align}
&J_c^{ext}(\bm u,\bm v,r^d,x(k))=\sum_{s=1}^2 J_1(\bm u_s,r_s^d,x_s(k)\vert \bm v_s) \label{costbis} \\
&:=\sum_{s=1}^2\Bigl[\sum_{i=1}^N\left(\dfrac{i}{N}\right)^q\left[\left\|h_s(k+i)-r_s^d\right\|_{Q_c^{(s)}}^2+\left\|U_s^0+u_s(k+i)\right\|_{R_c^{(s)}}^2\right]
\Bigr]
\end{align} 
where $h_s(k+\cdot)$ represents the nominal output profile defined by (\ref{shortys}):
\begin{equation}
h_s(\sigma):=h_s\Bigl(\bm X_s(\sigma,\bm u_s,x_s(k)\vert \bm v_s),\bm u_s(\sigma),\bm v_s(\sigma),0\Bigr) \label{defdehs} 
\end{equation} 
provided that the coupling signal profiles $\bm v=(\bm v_1,\bm v_2)$ satisfy the coherence constraints (\ref{coherenceeq1})-(\ref{coherenceeq2}) stated in the previous section:
\begin{align}
\bm v_1=\bm g_1(\bm u_2,x_2(k),\bm v_2) \label{coherenceeq1bis}\\ 
\bm v_2=\bm g_2(\bm u_1,x_1(k),\bm v_1) \label{coherenceeq2bis}
\end{align} 
Note that in the above cost (\ref{costbis}), the two individual costs are now conceptually decoupled for any given choice of the coupling signal profiles $\bm v_1$ and $\bm v_2$. The coupling appears now in the equality constraints (\ref{coherenceeq1bis})-(\ref{coherenceeq2bis}).  

Consequently, the central optimization problem can be rewritten in the following form:
\begin{align}
\min_{\bm u_1,\bm u_2,\bm v_1,\bm v_2} \sum_{s=1}^2 J_s(\bm u_s,r_s^d,x_s(k)\vert \bm v_s)\quad \mbox{\rm under} \quad \left\{\begin{array}{l}
\bm v_1=\bm g_1(\bm u_2,x_2(k),\bm v_2) \\ 
\bm v_2=\bm g_2(\bm u_1,x_1(k),\bm v_1)
\end{array}\right. \label{centralizedbis} 
\end{align} 
The last form (\ref{centralizedbis}) of the central optimization problem is already amenable to hierarchical distribution although it is not the one that will be finally used. Indeed, the coordinator can begin with some initial guesses:
$$\bm v_1^{(0)}, \bm v_2^{(0)}$$
of the coupling variables profiles. It sends $\bm v_s^{(0)}$ to subsystem $S_s$. Each subsystem is now able to compute its own contribution to the central cost, namely: $$J_s(\bm u_s,r_s^d,x_s(k)\vert\bm v_s^{(0)})$$ for any candidate control profile $\bm u_s$. Therefore, each subsystem can solve its own optimization problem and get its own optimal profile and cost:
\begin{equation}
\bm u_s^*(r_s^d,x_s(k)\vert\bm v_s^{(0)})\quad,\quad J_s^*(r_s^d,x_s(k)\vert\bm v_s^{(0)}) \label{ustar} 
\end{equation} 
Moreover, each subsystem sends what would be the corresponding coupling profile it would apply in this case and {\bf send it to the coordinator}:
\begin{align}
\mbox{\rm subsystem $S_1$ send to the coordinator}\quad \rightarrow \quad \hat{\bm v}_2^{(1)}:=\bm g_2(\bm u_1^*,x_1(k),\bm v_1^{(0)}) \label{reph1} \\
\mbox{\rm subsystem $S_2$ send to the coordinator}\quad \rightarrow \quad \hat{\bm v}_1^{(1)}:=\bm g_1(\bm u_2^*,x_2(k),\bm v_2^{(0)}) \label{reph2} 
\end{align} 
where $\bm u_s^*:=\bm u_s^*(r_s^d,x_s(k)\vert\bm v_s^{(0)})$. 

Having the new estimations $\hat{\bm v}_s^{(1)}$, the coordinator elaborates a {\em filtered} version of the coupling profiles:
\begin{equation}
\bm v_s^{(1)}:=(1-\beta)\bm v_s^{(0)}+\beta\hat{\bm v}_s^{(1)}
\end{equation} 
and therefore, a fixed-point round of iterations can take place. 

The reason why the fixed-point iteration will not be defined as explained above lies in the fact that the centralized cost formulation, when used to built up an MPC feedback design, does not necessarily lead to a stable behavior of the system. Indeed, as it is noticed earlier, the stability of a finite horizon MPC schemes requires some conditions on the formulation of the cost function (penalizing the whole state, zeros cost at the desired state, positive definite penalty on the control action and so on). Now modifying the central cost in order to meet these requirements may lead to an irrelevant cost from the economic/performance point of view. 

Instead, stabilizing formulations are used in the local level in order to enforce stability while only the local set-points $r_s$ sent by the coordinator to the subsystems are kept as decision variables for the coordinator. This is explained in the remainder of the current section. 

More precisely, for each subsystem $s$, given a set-point $r_s$ (not necessarily equal to the central set-point $r_s^d$) together with a current estimation of the coupling profile\footnote{$\sigma=0$ was used in the previous discussion} $\bm v^{(\sigma)}$, a local, standard well defined and stability inducing cost function is defined:
\begin{equation}
J_s^{MPC}(\bm u_s,r_s,x_s(k)\vert \bm v_s^{(\sigma)}):=\sum_{i=1}^N\|x_s(k+i)-x_s^d(r_s))\|^2_{Q_s}+\|\bm u_s(k+i)-u_s^d(r_s))\|^2_{R_s} \label{localMPC} 
\end{equation} 
where 
\begin{itemize}
\item[$\checkmark$] $(x_s^d(r_s),u_s^d(r_s))$ is the steady pair that is compatible with the output set-point $r_s$.\\
\item[$\checkmark$] $x_s(k+i):=\bm X_s(k+i,\bm u_s,x_s(k)\vert \bm v_s^{(\sigma)})$ is the nominal predicted state at  instant $k+i$ given the initial state $x_s(k)$, the control profile $\bm u_s$ and the presumed coupling profile $\bm v_s^{(\sigma)}$.
\end{itemize} 
The optimal solution of this well-posed optimization problem in the decision variables $\bm u_s$ is denoted by $$\bm u_s^{opt}(r_s,x_s(k)\vert \bm v_s^{(\sigma)})$$
and it is now this sequence that is used to construct the fixed point iteration  instead of $\bm u^*(r_s^d,x_s(k)\vert \bm v_s^{(\sigma)})$ as it is suggested earlier [see (\ref{ustar}) and the development that followed]. 

Therefore, rephrasing (\ref{reph1})-(\ref{reph2}) with $\bm u_s^{opt}$ instead of $\bm u_s^*$ and $\bm v_s^{(\sigma)}$ instead of $\bm v_s^{(0)}$ leads to:
\begin{align}
\mbox{\rm $S_1$ send to the coordinator}\quad \rightarrow \quad \hat{\bm v}_2^{(\sigma+1)}:=\bm g_2(\bm u_1^{opt}(r_1),x_1(k),\bm v_1^{(\sigma)}) \label{reph1} \\
\mbox{\rm $S_2$ send to the coordinator}\quad \rightarrow \quad \hat{\bm v}_1^{(\sigma+1)}:=\bm g_1(\bm u_2^{opt}(r_2),x_2(k),\bm v_2^{(\sigma)}) \label{reph2} 
\end{align} 
where $\bm u_s^{opt}(r_s):=\bm u_s^{opt}(r_s,x_s(k)\vert\bm v_s^{(\sigma)})$. 

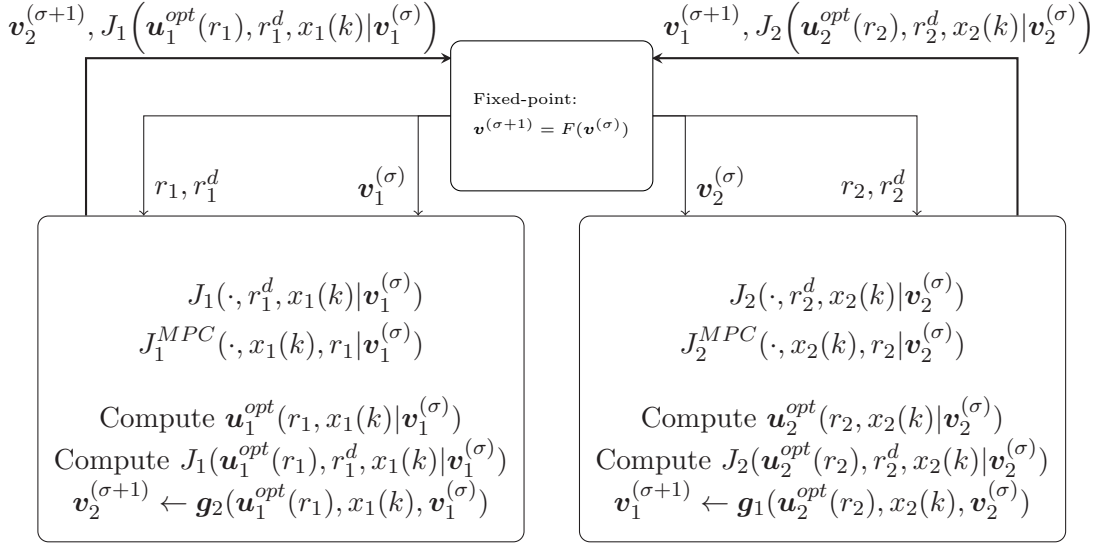
\begin{figure*}
\begin{center}
\vskip -1cm 
\begin{tikzpicture}
\node[rounded corners,draw,Orange,inner xsep=3mm,inner ysep=8mm] at(0,0)(C){\footnotesize Coordinator};
\node[rounded corners,draw,Black,inner xsep=1mm,inner ysep=3mm] at(-3.6,-3.5)(S1){
\begin{minipage}{0.45\textwidth}
\footnotesize
\begin{align*}
J_1(\cdot,r_1^d,x_1(k)\vert \bm v_1^{(\sigma)})\\
J_1^{MPC}(\cdot,x_1(k),r_1\vert \bm v_1^{(\sigma)})
\end{align*} 
\begin{center}
Compute $\bm u^{opt}_1(r_1,x_1(k)\vert \bm v_1^{(\sigma)})$\\
Compute $J_1(\bm u^{opt}_1(r_1),r_1^d,x_1(k)\vert \bm v_1^{(\sigma)})$\\
$\bm v_2^{(\sigma+1)}\leftarrow \bm g_2(\bm u_1^{opt}(r_1),x_1(k),\bm v_1^{(\sigma)})$
\end{center} 
\end{minipage} 
};
\node[rounded corners,draw,Black,inner xsep=1mm,inner ysep=3mm] at(3.6,-3.5)(S2){
\begin{minipage}{0.45\textwidth}
\footnotesize
\begin{align*}
J_2(\cdot,r_2^d,x_2(k)\vert \bm v_2^{(\sigma)})\\
J_2^{MPC}(\cdot,x_2(k),r_2\vert \bm v_2^{(\sigma)})
\end{align*} 
\begin{center}
Compute $\bm u^{opt}_2(r_2,x_2(k)\vert \bm v_2^{(\sigma)})$\\
Compute $J_2(\bm u^{opt}_2(r_2),r_2^d,x_2(k)\vert \bm v_2^{(\sigma)})$\\
$\bm v_1^{(\sigma+1)}\leftarrow \bm g_1(\bm u_2^{opt}(r_2),x_2(k),\bm v_2^{(\sigma)})$
\end{center} 

\end{minipage} 
};
\draw[->,Black,thin] (C.west) -| (S1.130) node[above=4mm,right]{\footnotesize$r_1,r^d_1$};
\draw[->,Black,thin] (C.west) -| (S1.50) node[above=4mm,left]{\footnotesize$\bm v_1^{(\sigma)}$};
\draw[->,Black,thin] (C.east) -| (S2.60) node[above=4mm,left]{\footnotesize$r_2,r^d_2$};
\draw[->,Black,thin] (C.east) -| (S2.130) node[above=4mm,right]{\footnotesize$\bm v_2^{(\sigma)}$};
\draw[->,thick,>=stealth,Black] (S2.40) |- (C.30) node[above=4mm,right]{\footnotesize $\bm  v_1^{(\sigma+1)},J_2\Bigl(\bm u_2^{opt}(r_2),r_2^d,x_2(k)\vert \bm v_2^{(\sigma)}\Bigr)
$};
\draw[->,thick,>=stealth,Black] (S1.140) |- (C.150) node[above=4mm,left]{\footnotesize $\bm v_2^{(\sigma+1)},J_1\Bigl(\bm u_1^{opt}(r_1),r_1^d,x_1(k)\vert \bm v_1^{(\sigma)}\Bigr)$};
\begin{scope}[overlay]
\node[rounded corners,draw,fill=white,inner xsep=3mm,inner ysep=7mm] at(0,0)(C){
\begin{minipage}{0.15\textwidth}
\relsize{-3} \color{black} Fixed-point:\vskip 0.1cm 
$\bm v^{(\sigma+1)}=F(\bm v^{(\sigma)})$
\end{minipage} 
};
\end{scope}
\end{tikzpicture}
\end{center} 
\caption{Schematic of the fixed-point iterations that takes place at instant $k$, for a {\em frozen} states $x_1(k)$, $x_2(k)$, and frozen set-points $r_s$ and $r_s^d$.} \label{FP_iteration} 
\end{figure*}

Having the new estimations $\hat{\bm v}_s^{(\sigma+1)}$, the coordinator elaborates a {\em filtered} version of the coupling profiles:
\begin{equation}
\bm v_s^{(\sigma+1)}:=(1-\beta)\bm v_s^{(\sigma)}+\beta\hat{\bm v}_s^{(\sigma+1)} \label{FFPit} 
\end{equation} 
and therefore, a fixed point-like round of iterations can take place. The whole scheme is sketched in Figure \ref{FP_iteration}.

Note that in order for the coordinator to compute the value of the central cost function for a given pair $(r_1,r_2)$ of local set-points, each subsystem evaluates its contribution to the central cost (depending on the central set-points $r_s^d$), {\bf at the optimal} profile $\bm u_s^{opt}(r_s)$, namely:
\begin{equation}
J_s(\bm u_s^{opt}(r_s),r_s^d,x_s(k)\vert \bm v_s^{(\sigma)})\quad \mbox{\rm where}\   \bm u_s^{opt}(r_s):=\bm u_s^{opt}(r_s,x_s(k)\vert \bm v_s^{(\sigma)})
\end{equation} 
and sends it to the coordinator. Upon receiving these evaluations, the coordinator can compute the estimate the value of the central cost for the set-point $r:=(r_1,r_2)$:
\begin{equation}
J(r\vert r^d,\bm v^{(\sigma)}):=\sum_{s=1}^2\underbrace{J_s\Bigl(\bm u_s^{opt}(r_s),r_s^d,x_s(k)\vert \bm v_s^{(\sigma)}}_{\mbox{\rm sent by subsystem $S_s$} }\Bigr) \label{jhjhjh45} 
\end{equation} 
\begin{rem}
It is important to underline here that the estimation so obtained by the coordinator does not involve the knowledge of the states $x_s(k)$ nor that of the optimal sequence of control $\bm u^{(opt})$, these quantity are only known locally in the subsystem which sends the resulting corresponding value of $J_s$. 
\end{rem}

It is important to underline that the above estimation is irrelevant {\bf unless the fixed point iteration converges} toward some fixed point $\bm v^{(\infty)}$ since the coherence constraint is not fulfilled otherwise. Said equivalently, if one has:
\begin{equation}
\lim_{\sigma\rightarrow\infty} \bm v^{(\sigma)}=\bm v^{(\infty)}
\end{equation} 
then the quadruplet:
\begin{equation}
\Bigl(\bm u_1^{opt}(r_1,x_1(k)\vert \bm v_1^{(\infty)}), \bm u_1^{opt}(r_2,x_2(k)\vert \bm v_2^{(\infty)}), \bm v_1^{(\infty)}, \bm v_2^{(\infty)}\Bigr)
\end{equation}  
is an admissible sub-optimal solution to the constrained central optimization problem (\ref{centralizedbis}). Therefore, only in this case, one can consider that $J(r\vert r^d,\bm v^{(\infty)})$ is the true value of the central cost when the control profiles $\bm u_s^{opt}$ are applied by the subsystems. 

In order to smoothly introduce the remaining part of the control strategy, the analysis of the conditions under which the above defined fixed-point iteration converges is delayed to Section \ref{subsectionCA}. Meanwhile, the next section explains how to built the overall hierarchical feedback in case this convergence unconditionally holds. 
\subsection{Closing the loop} \label{subsectionclosetheloop} 
For each value of the auxiliary set-point $r=(r_1,r_2)$ the coordinator settles to define the fixed-point round of negotiation with the subsystems, the coordinator obtains after convergence the value of the central cost for this set-point vector $r\in \mathbb{R}^{n_r}$, namely $$J(r\vert r^d, \bm v^{(\infty)}(r))$$
which is defined by (\ref{jhjhjh45}). Now since the above function is quadratic in $r$, the coordinator needs to evaluate its value over some grid including more than or equal to $(n_r+1)(n_r+2)/2$ different nodes in order to reconstruct the central cost as a quadratic function of $r$, namely:
\begin{equation}
\hat J(r)=\dfrac{1}{2}r^TQr+f^Tr+c
\end{equation} 
where $Q\in \mathbb{R}^{n_r\times n_r}$, $f\in \mathbb{R}^{n_r}$ and $c\in \mathbb{R}$ depends obviously on $r^d$ and $x(k)$ in a way that is analytically unknown to the coordinator. 

This can obviously be done by taking a regular grid around $r^d$ such as the one defined by:
\begin{align}
r_i^{(j)}&:=r_i^d+g^{(j)}_i\cdot \Delta\quad& (i,j)\in \{1,\dots,n_r\}\times\{1,\dots,m^{n_r}\} \\
g^{(j)}&\in \left[\mathcal N_m\right]^{n_r}\quad &N_m:=\{-1,\dots,+1\}\in \mathbb{R}^{m}
\end{align} 
for some $\Delta>0$ and some integer $m$ such that $m^{n_r}\ge (n_r+1)(n_r+2)/2$. The coefficients $Q$, $f$ and $c$ of the quadratic form can therefore be computed by solving the least squares problem:
\begin{equation}
\min_{Q,f,c} \sum_{j=1}^{m^{n_r}}\left\vert \hat J(r^{j})-[\frac{1}{2}\|r^{(j)}\|_Q^2+f^Tr^{(j)}+c]\right\vert^2
\end{equation} 
Once $Q$, $f$ are available, the optimal auxiliary set-point can be computed according to:
\begin{equation}
r^{opt}(k):=-Q^{-1}f \label{solropt} 
\end{equation} 
This optimal set-point is therefore sent to the subsystems with an {\em end-of-iterations} flag which is interpreted by the subsystem as the definitive optimal set-point at instant $k$. Based on this information and in accordance with the receding-horizon principle, the first action in the corresponding optimal sequences, namely
\begin{equation}
u_s(k):= \begin{bmatrix}
\mathbb I_{n^u_s},\mathbb O_{n^u_s},\dots,\mathbb O_{n^u_s}
\end{bmatrix}  \bm u_s^{opt}(r^{opt}_s(k),x_s(k)\vert \bm v_s^{(\infty)})
\end{equation}  
is applied by subsystem $S_s$ during the sampling period $[k,k+1]$. 

Note that this implicitly assumes that the computation time for the $m^{n_r}$ iterations is negligible when compared to the sampling period of the control loop. Note also that the scheme should include an anticipative action in the sense that the computation of $u_s(k)$ should be done during the previous sampling period $[k-1,k]$ based on each subsystem estimation of its future state at instant $k$. These are standard implementation tricks that have been skipped here in order to focus on the main message. 

This completely define the hierarchical control algorithm. In the next section, the analysis of the convergence of the fixed point iteration is done. 
\subsection{Convergence of the fixed-point iteration} \label{subsectionCA} 
In order to analyze the convergence of the fixed point iteration, we begin by establishing the dynamics that governs the successive iterates $\bm v^{(\sigma)}$. To do this, note that at each iteration, subsystem $S_s$ solves a finite horizon unconstrained quadratic optimization problem in which the only exogenous information is represented by the current state $x_s(k)$, the auxiliary set-point $r_s$ and the exogenous coupling signal $\bm v_s^{(\sigma)}$, therefore, the optimal control profile $\bm u^{opt}_s$ computed by subsystem $S_s$ takes the following form:
\begin{align}
\bm u^{opt}_1:=[K_1^{(x)}]x_1(k)+[K_1^{(r)}]r_1+[K_1^{(v)}]\bm v_1^{(\sigma)} \label{K1} \\
\bm u^{opt}_1:=[K_1^{(x)}]x_2(k)+[K_2^{(r)}]r_2+[K_2^{(v)}]\bm v_2^{(\sigma)} \label{K2} 
\end{align} 
with straightforward definition of the gain matrices. 

On the other hand, the updated value $\hat{\bm v}_1^{(\sigma+1)}$ [resp. $\hat{\bm v}_1^{(\sigma+1)}$] is nothing but output of a linear system with initial state $x_1(k)$ [resp. $x_2(k)$], input profiles given by (\ref{K1}) [resp. (\ref{K2})] and affected by coupling signal $\bm v_1^{(\sigma)}$ [resp. $\bm v_2^{(\sigma)}$]. This can be written as follows:
\begin{align}
\hat{\bm v}_1^{(\sigma+1)}:=[M_2^{(x)}]x_2(k)+[M_2^{(u)}]\bm u_2^{opt}+[M_2^{(v)}]\bm v_2^{(\sigma)} \label{M1}\\ 
\hat{\bm v}_2^{(\sigma+1)}:=[M_1^{(x)}]x_1(k)+[M_1^{(u)}]\bm u_1^{opt}+[M_1^{(v)}]\bm v_1^{(\sigma)} \label{M2}
\end{align} 
withstraightforward definition of the gain matrices. Combining (\ref{K1})-(\ref{M2}) obviously leads to:
\begin{align}
\hat{\bm v}^{(\sigma+1)}=& \left[\bar M^{(v)}\right]\bm v^{(\sigma)}  +\left[\bar M^{(r)}\right]r+\left[\bar M^{(x)}\right]x(k) \label{justavant} 
\end{align} 
where 
\begin{align}
\left[\bar M^{(v)}\right]&:=\begin{bmatrix}
\mathbb O& M_2^{(v)}+M_2^{(u)}K_2^{(v)}\cr M_1^{(v)}+M_1^{(u)}K_1^{(v)} & \mathbb O
\end{bmatrix}=: \begin{bmatrix}
\mathbb O & \bar M_2^{(v)}\cr \bar M_1^{(v)} & \mathbb O
\end{bmatrix}   \\
\left[\bar M^{(r)}\right]&:=\begin{bmatrix}
\mathbb O& M_2^{(r)}+M_2^{(u)}K_2^{(r)}\cr M_1^{(r)}+M_1^{(u)}K_1^{(r)} & \mathbb O
\end{bmatrix} \\
\left[\bar M^{(x)}\right]&:=\begin{bmatrix}
\mathbb O& M_2^{(x)}+M_2^{(u)}K_2^{(x)}\cr M_1^{(x)}+M_1^{(u)}K_1^{(x)} & \mathbb O
\end{bmatrix} 
\end{align} 
and finally, using (\ref{justavant}) in the updating rule (\ref{FFPit}) one obtains the dynamics that governs the fixed point iteration:
\begin{equation}
\bm v^{(\sigma+1)}=\underbrace{\left[(1-\beta)\mathbb I+\beta \bar M^{(v)}\right]}_{Z(\beta)}\bm v^{(\sigma)}+\beta\Bigl[\left[\bar M^{(r)}\right]r+\left[\bar M^{(x)}\right]x(k)\Bigr] \label{defdeZbeta} 
\end{equation} 
This clearly indicates that the convergence of the fixed-point iteration is conditioned by the spectrum radius of the matrix $Z(\beta)$ defined above. More precisely:
\begin{equation}
\Bigl\{\mbox{\rm The fixed-point iteration converges}\Bigr\} \Leftrightarrow  \Bigl\{\rho(Z(\beta))<1\Bigr\} \label{convvvv} 
\end{equation} 
where $\rho(Z)$ is the spectrum radius of the matrix:
\begin{equation}
\rho(Z(\beta)):= \max_{i}\vert \lambda_i(Z(\beta))\vert
\end{equation} 
The above arguments enables to state the following result:
\begin{center}
\begin{minipage}{0.8\textwidth}
{\bf Fixed-point Convergence certification}\\ \ \\
The matrices $\{\bar M_s^{(v)}\}_{s\in \{1,2\}}$ summarize the static information the coordinator needs to have in order to be able to check and to certify the convergence (if any) of the fixed-point iteration that is in the heart of the proposed hierarchical framework. Having these matrices, the coordinator might appropriately choose the filtering parameter in order to enhance the convergence. 
\end{minipage} 
\end{center} 
\begin{rem}
Note that the matrices $\bar M_s^{(v)}$ represent a high level condensed information that does not depend on the state of subsystem $S_s$, neither it depends on the number of actuators or the precise control law settings that are used in the local level.  
\end{rem}
\begin{rem}
The filtering law (\ref{FFPit}) is obviously over simplified. More elaborated filters can be used in case the simple rule does not allow convergence for any possible value of $\beta\in (0,1)$. This is not elaborated here for the sake of simplicity and since the simple rule is sufficient for the process under study. 
\end{rem}
\section{Simulation results} \label{secresult} 
In this section, the hierarchical MPC framework proposed in the previous section is applied to address the hierarchical control problem of the cryogenic refrigerator. More precisely, the following validation topics are successively addressed. 
\begin{enumerate}
\item First of all, the convergence of the fixed-point iteration is analyzed for different values of the filtering parameter $\beta$. \\
\item Then the by-default regulation scenario is simulated and the performance of the closed-loop system is shown. \\
\item Two typical scenarios of set-point changes on the outputs Ltb$_{131}$ and Ttb$_{108}$ are successively simulated under the proposed control framework. \\
\item The behavior of the fully decentralized control (assuming zero coupling signals by each local controller) is shown in order to underline the benefit from the use of the hierarchical structure. 
\end{enumerate} 
Before addressing the issues enumerated above, some parameters that are used throughout the simulations are first described in the following section. 
\subsection{The simulation parameters}
The sampling time for the simulation is taken equal to $\tau=5$ seconds. This is the currently used sampling time at CEA/INAC/SBT. Moreover, it is the one that has been used in all previous studies \cite{Bonne2013,Bonne2014JPC,Bonne2015}. The prediction horizon is taken equal to $N_p\tau$ where $N_p=100$. This corresponds to roughly 8 minutes. 

The local MPC controllers are based on the local cost function (\ref{localMPC}) in which the following trivial weighting matrices are used:
\begin{align}
Q_1&:=10^6\times \mathbb I_{10\times 10}\ ;\ R_1= \begin{bmatrix}
1&0\cr 0&10
\end{bmatrix}  \ ;\ Q_2=\mathbb I_{14\times 14}\ ;\ R_2=10^2
\end{align} 
Regarding the weighting matrices used in the central cost, the exponent $q=20$ is used in the time-varying weighting settings; the control weighting matrices were taking systematically equal to $R_c^{(s)}=0$, $s=1,2$ in order to focus on the regulation performance. Regarding the output regulation weighting matrices $Q_c^{(s)}$, three different settings are used depending on the control mode, namely:
\begin{enumerate}
\item In the by default disturbance rejection mode, the following settings is used:
\begin{equation}
Q_c^{(1)}= \begin{bmatrix}
10^2&0\cr 0& 10^6
\end{bmatrix}\quad ;\quad Q_c^{(2)}=1  
\end{equation} 
\item When the second mode is activated in order to steer the level of the helium bath to some desired value, the following setting is used:
\begin{equation}
Q_c^{(1)}= \begin{bmatrix}
10^6&0\cr 0& 1
\end{bmatrix}\quad ;\quad Q_c^{(2)}=1  
\end{equation}
\item Finally when the second mode is activated in order to steer the temperature Ttb$_{108}$ to some desired value, the following settings is used:
\begin{equation}
Q_c^{(1)}= \begin{bmatrix}
1&0\cr 0& 10^6
\end{bmatrix}\quad ;\quad Q_c^{(2)}=1  
\end{equation}  
\end{enumerate} 
Regarding the quantities involved in the identification of the quadratic cost by the coordinator (see Section \ref{subsectionclosetheloop}), we have $n_r=3$ and $m=3$ and $\Delta=1$ are used. 
\subsection{Fixed-point convergence analysis}
\begin{figure}
\begin{center}
\input{spectrum.tex}
\end{center} 
\caption{Evolution of the spectrum radius of the matrix $Z(\beta)$ [see (\ref{defdeZbeta})] governing the stability of the fixed point iteration as a function of the filtering coefficient $\beta$.} \label{spectrum} 
\end{figure}
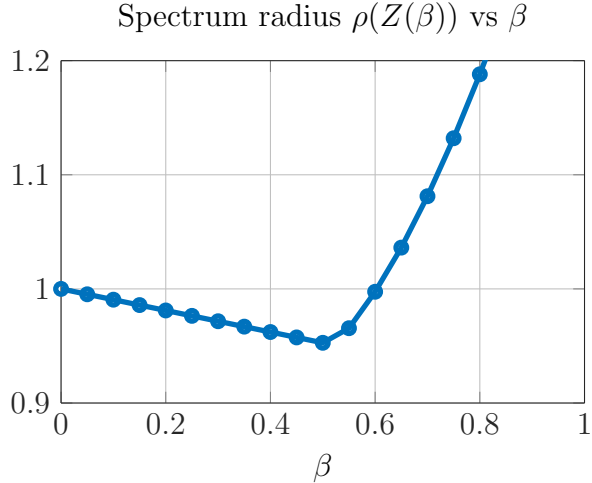

Recall that the convergence condition (\ref{convvvv}) for the fixed point iteration does not depends on the settings of the central cost. It rather depends on the local settings and the filtering parameter $\beta$. Figure \ref{spectrum} shows the evolution of the spectrum radius $\rho(Z(\beta))$ involved in the stability condition (\ref{convvvv}) as a function of the filtering coefficient $\beta$. This figure clearly shows that the choice $\beta=0.5$ is appropriate in terms of the contraction ratio. 

\begin{figure}
\begin{center}
\includegraphics[width=0.6\textwidth]{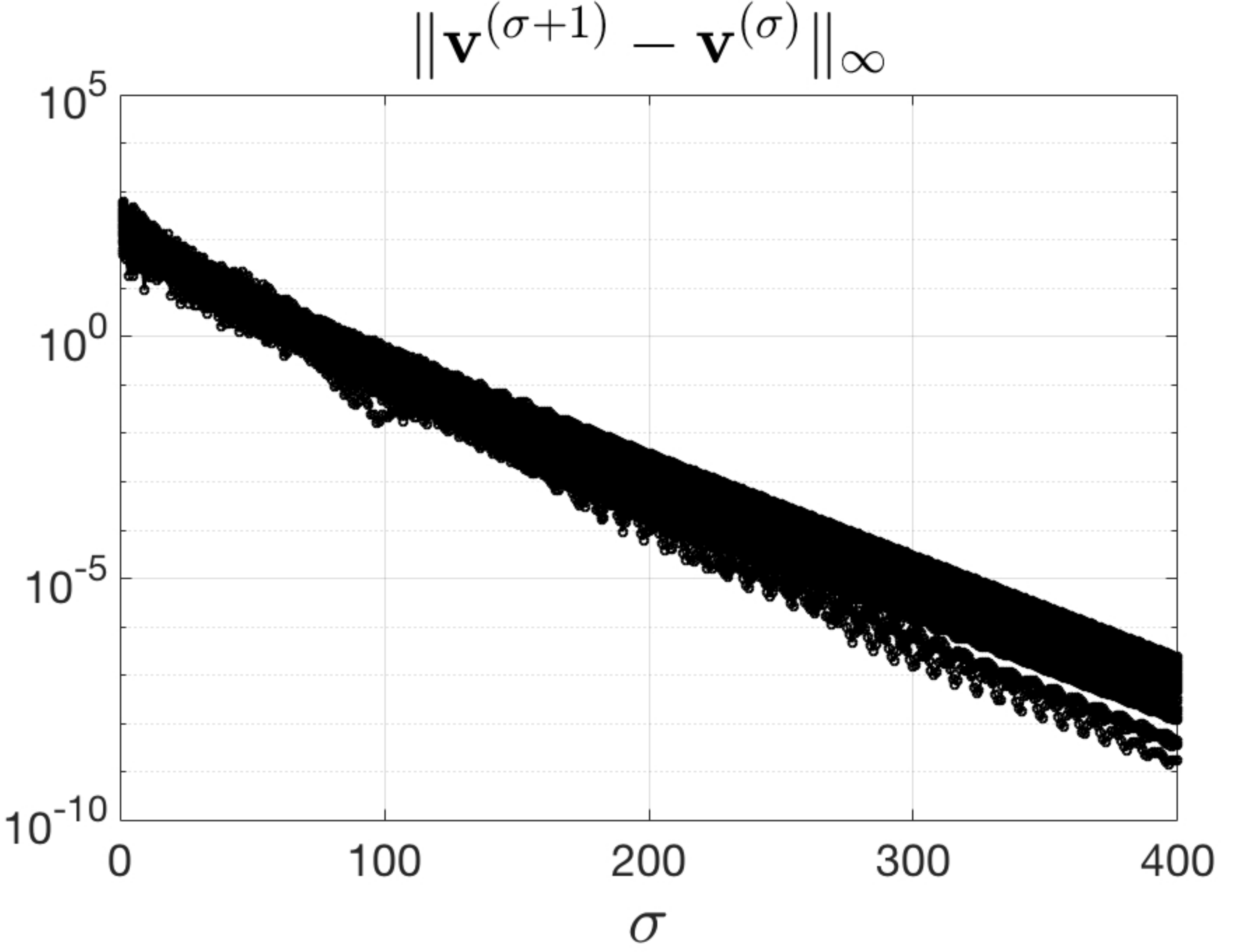} 
\end{center} 
\caption{Evolution of the maximum error between two successive coupling signal during the fixed point iterations. $50$ random initial values $\bm v^{(0)}$ are simulated.} \label{convergence} 
\end{figure}

Figure \ref{convergence} shows the evolution of the maximum error between two successive iterates during the fixed-point iterations. $50$ random initial values $\bm v^{(0)}$ are simulated. From this figure it comes out that $400$ iterations generally lead to an error less than $10^{-5}$ which is assumed to be sufficiently small to stop the iterations. Therefore, this number of iterations is systematically used in the sequel although a context-dependent  varying number can be adopted by monitoring the error between two successive iterates. The fact is that a sampling period of $5$ seconds is already largely higher than the time needed for $400$ iterations ($\approx 0.31$ sec) and hence, adopting a stopping criteria does not change the result for our application.

\subsection{Closed-loop simulations}

\begin{figure}[H]
\begin{center}
\includegraphics[width=0.95\textwidth]{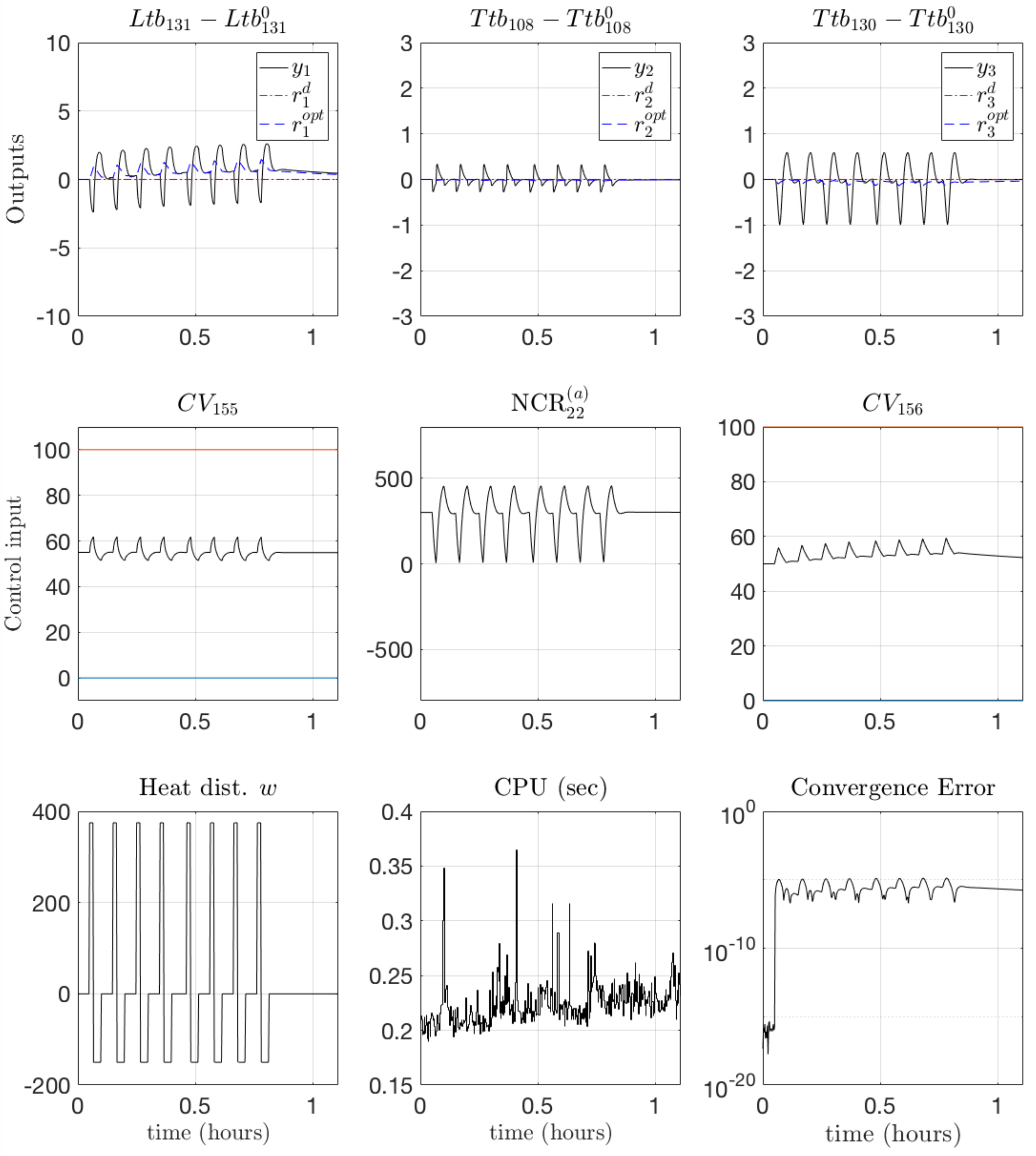} 
\end{center} 
\caption{Behavior of the closed-loop process under heat pulse disturbance } \label{exp1} 
\end{figure}

Figure \ref{exp1} shows the behavior of the closed-loop process in response to a periodic heat pulse disturbance. This disturbance emulates the heat disturbance that comes from the experimental facilities using the cooling power to cool down the supra-conducting circuits. The level of heat pulses used in this scenario is considered to be the sizing level for the experimental cryogenic refrigerator. Note that the computation time needed for the coordinator to deliver the optimal auxiliary set-points $r^{opt}$ never exceed 310 milliseconds. Moreover the convergence of the fixed-point iteration is assessed through the last plots where it can be observed that it is roughly around $10^{-5}$.

\begin{figure}[H]
\begin{center}
\includegraphics[width=0.95\textwidth]{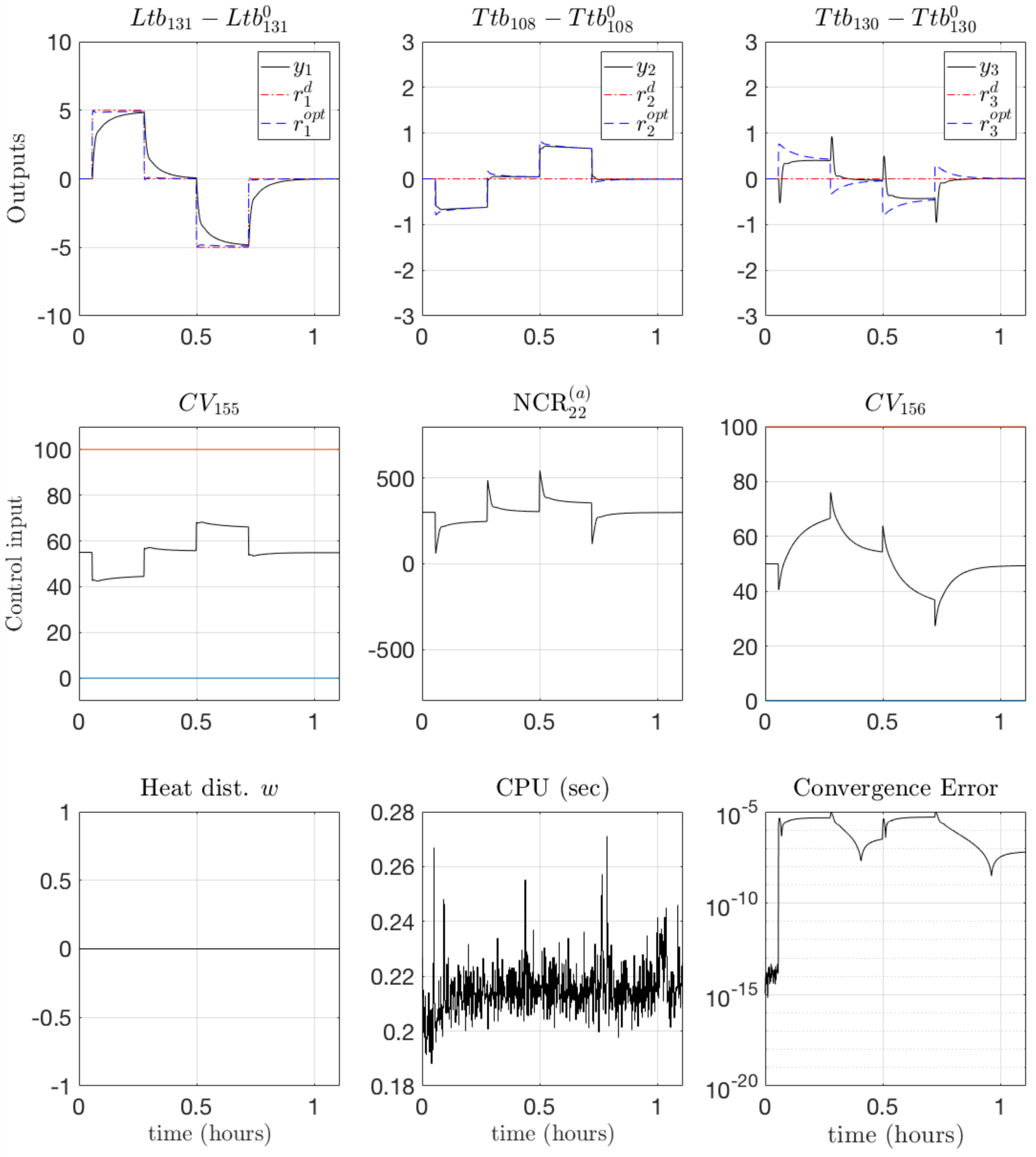} 
\end{center} 
\caption{Behavior of the closed-loop process in mode 2 in which the regulation of liquid level Ltb$_{131}$ is privileged. Note how the regulation of the temperature Ttb$_{108}$ is softened in order to perform the required task on the regulated level.} \label{exp2} 
\end{figure}

Figure \ref{exp2} shows the behavior of the closed-loop process when mode 2 is activated in order to steer the level of liquid in the helium bath to some desired values. Note how due to the specific settings of the weighting matrices of the central cost, the regulation of the temperature Ttb$_{108}$ is softened in favor of better regulation of the level. This can be shown by examining the difference between the central desired values $r^d$ (in axis-like red lines on the plots) and the auxiliary set-points $r^{opt}$  (in dashed blue lines on the plots). 

\begin{figure}[H]
\begin{center}
\includegraphics[width=0.95\textwidth]{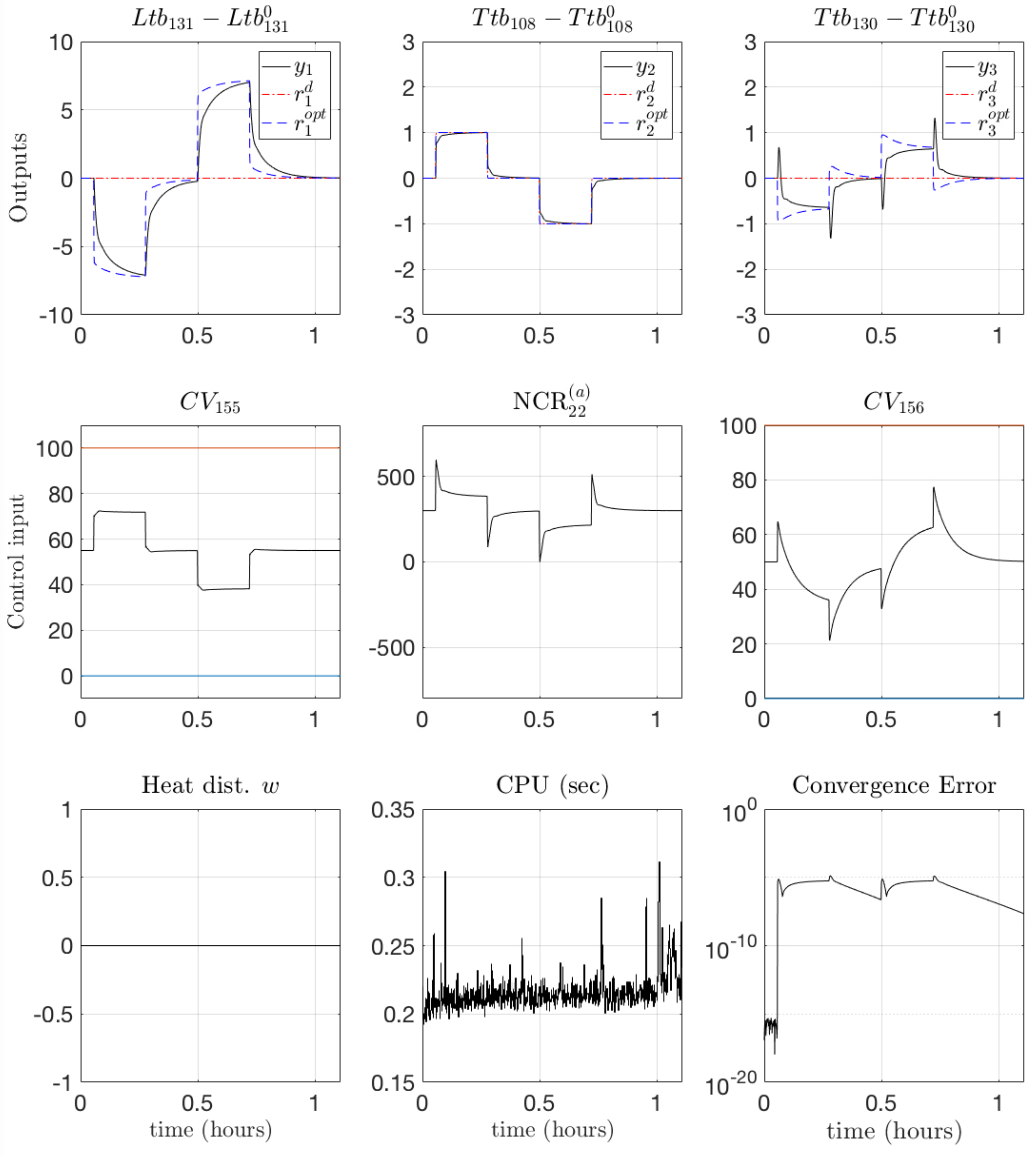} 
\end{center} 
\caption{Behavior of the closed-loop process in mode 2 in which the regulation of the variable Ttb$_{108}$ is privileged. Note how the regulation of the helium level in the bath is softened in order to perform the required task on the regulated temperature.} \label{exp3} 
\end{figure}

Figure \ref{exp3} shows the behavior of the closed-loop process when mode 2 is activated in order to steer the the temperature Ttb$_{108}$ of the JT valve to some desired values. Note how due to the specific settings of the weighting matrices of the central cost in this case, the regulation of the level  Ltb$_{131}$ in the helium bath of subsystem $S_1$ is softened in favor of better regulation of the JT temperature. This can be shown by examining the difference between the central desired values $r^d$ (in axis-like red lines on the plots) and the auxiliary set-points $r^{opt}$  (in dashed blue lines on the plots). 

\begin{figure}[H]
\begin{center}
\includegraphics[width=\textwidth]{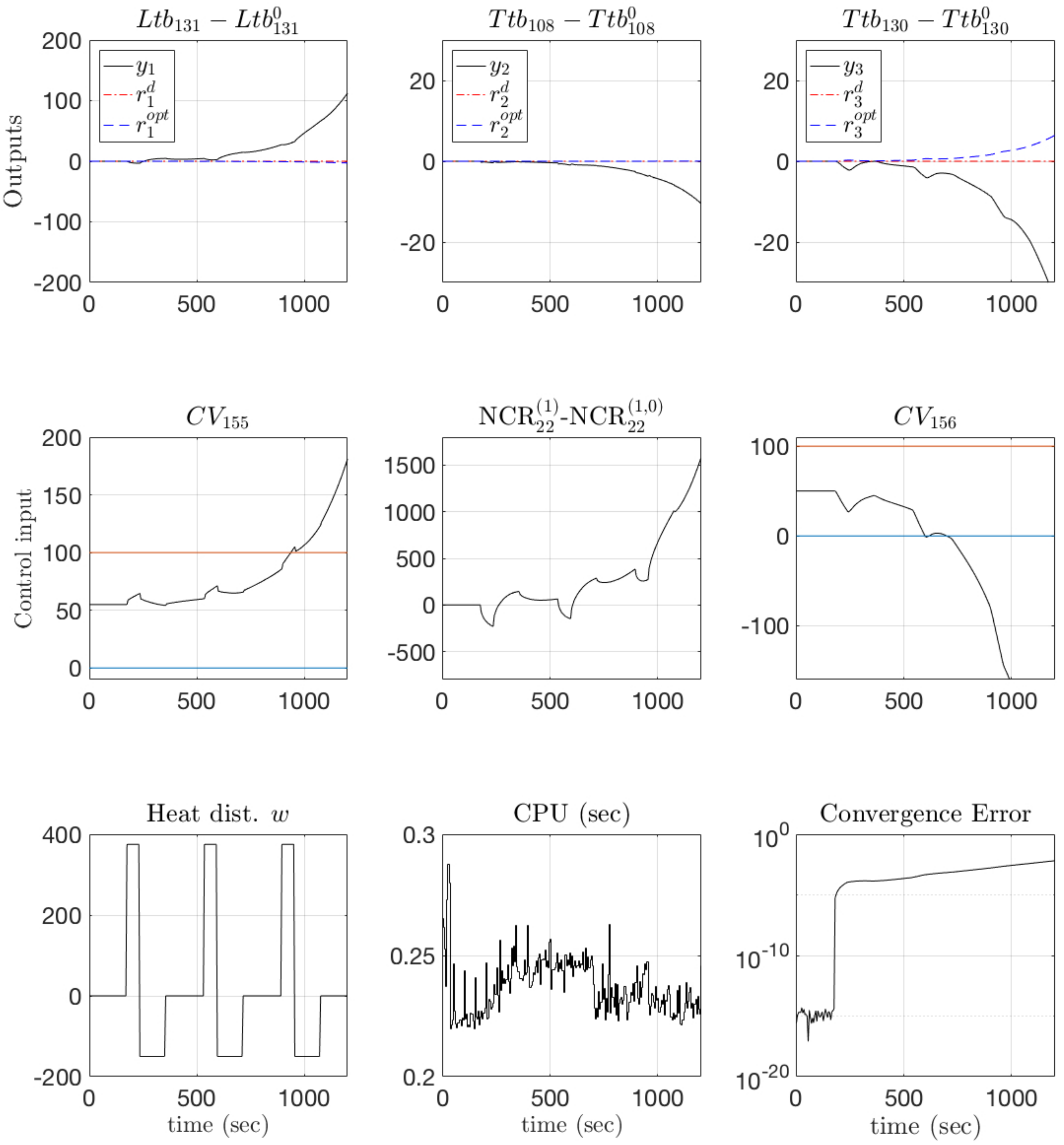} 
\end{center} 
\caption{Behavior of the {\bf fully-decentralized} closed-loop process under the heat pulse disturbance scenario. Here, each local controller considers that there are no coupling signal (all of them set to zeros). Note that the last two plots showing the computation time and the convergence error are not relevant in this case. } \label{exp4} 
\end{figure}

It is worth underlying that the number of iteration ($=400$) is probably over pessimistic and a precision of $10^{-5}$ on the fixed-point iterations error is not necessary. Simulations proving this fact are not produced here as the plots are roughly the same as the ones presented. This means that the computation time can be divided by $2$ without noticeable performance losses.

Finally, figure \ref{exp4} shows the behavior of the {\bf fully decentralized} controller, namely, when each local controller considers that the coupling signal are vanishing. This scenario underlines the destabilizing character of the coupling signals involved in the cryogenic refrigerator. It is worth emphasizing that in this context, the last two plots (cpu time and convergence error) are not relevant.

\section{Further discussion: handling operator handover} \label{secdiscussion} 
In industrial context, it is mandatory that the control architecture allows operator handover on the controlled process. These are predefined modes where the operator informs the control system that he (she) wishes to partially short-circuit the control logic in order to directly deliver his (her) own decision. 

\begin{figure}[H]
\begin{center}
\includegraphics[width=0.6\textwidth]{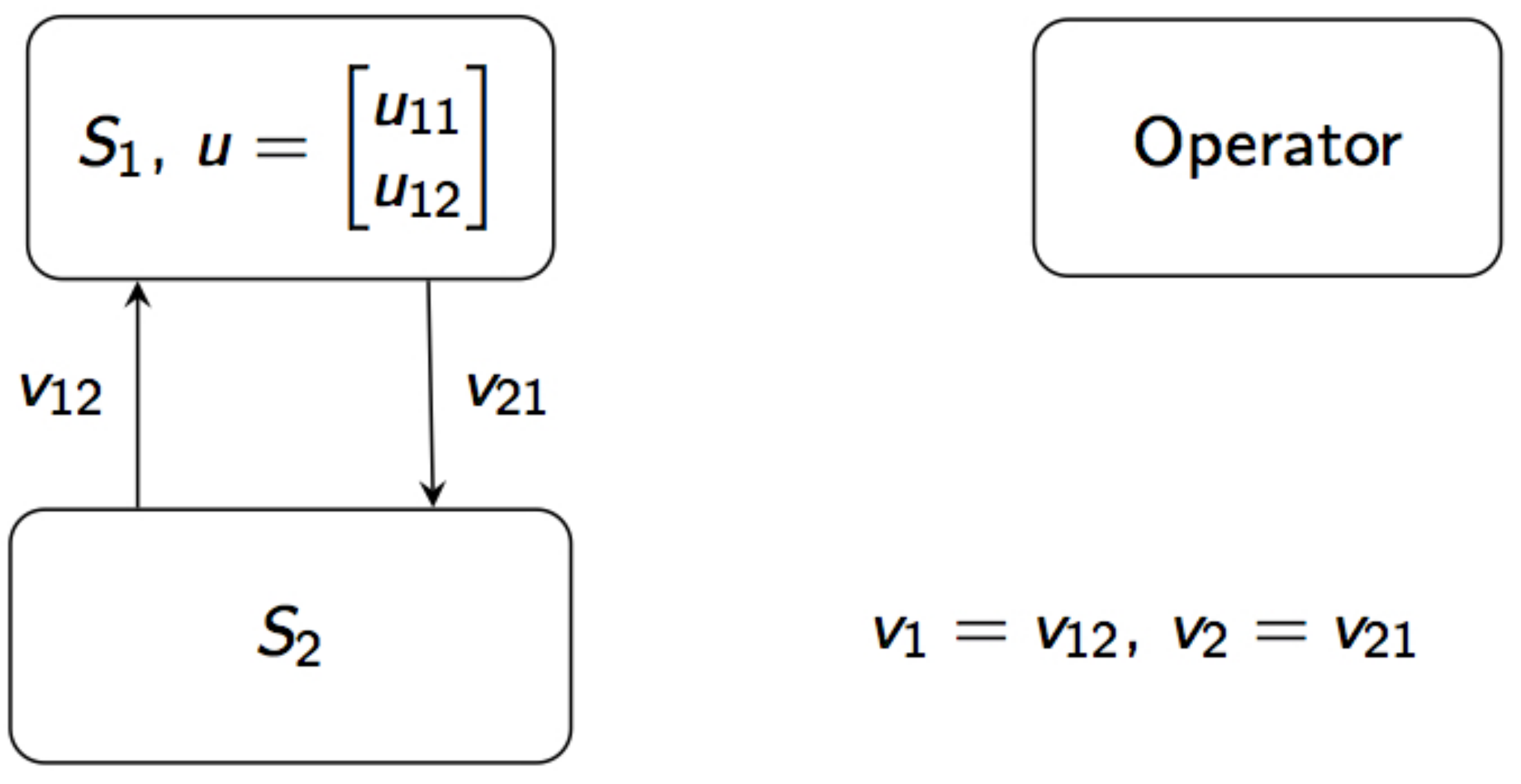} \\
(a) Before operator handover on $u_{11}$\vskip 1cm 
\includegraphics[width=0.6\textwidth]{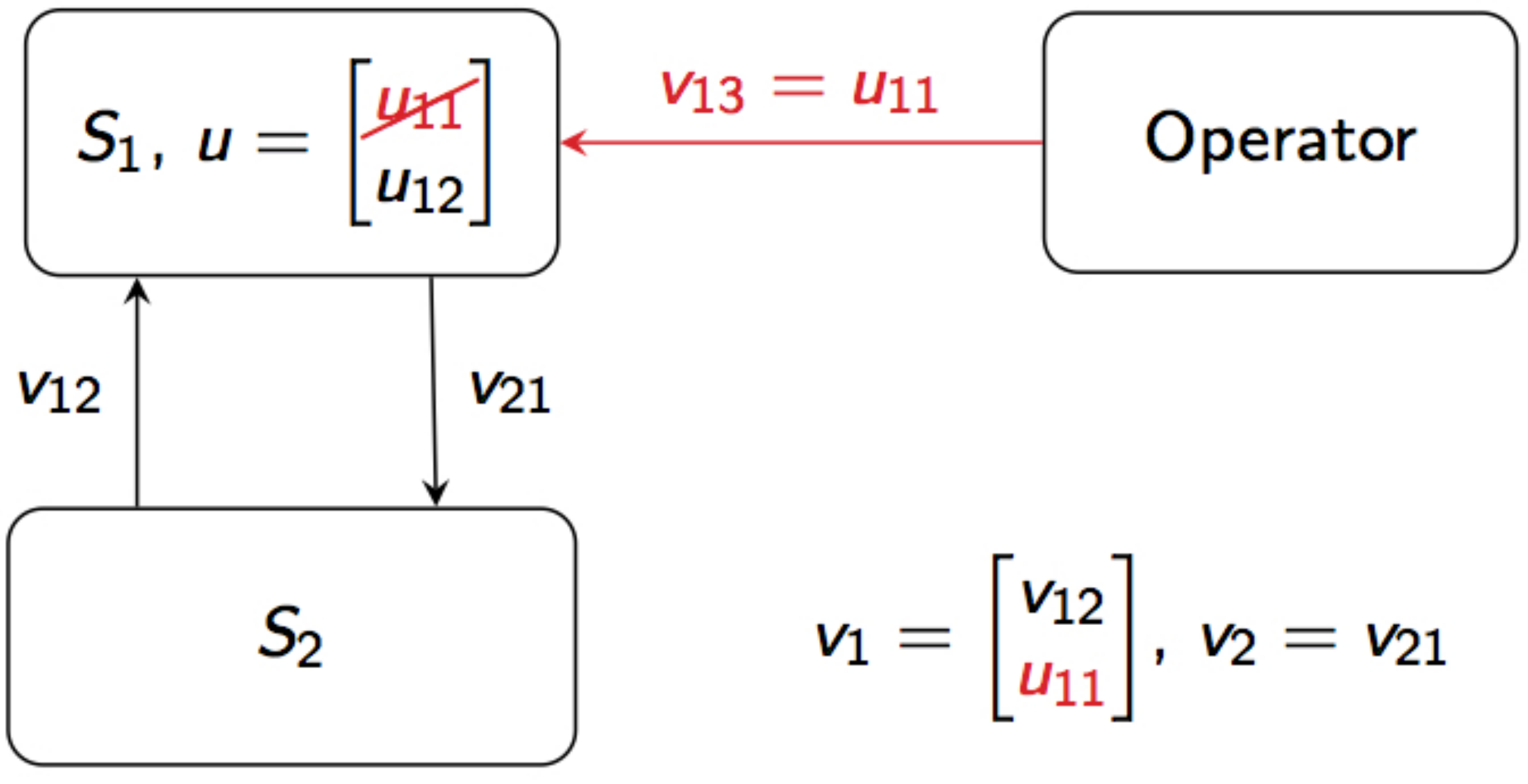}\\
(b) After operator handover on $u_{11}$
\end{center} 
\caption{Schematic view of the way low level operator handover can be handled.} \label{handover} 
\end{figure}

Depending on the level of the decision the operator wishes to deliver, two handover levels can be defined: 
\begin{enumerate}
\item In the first, the operator decides to deliver one (or more) auxiliary set-point, say $r_j$ instead of this set-point being computed by the coordinator of the hierarchical architecture. Handling this case is extremely easy as all the coordinator needs to do is to replace the unconstrained expression (\ref{solropt}) of the optimal $r^{opt}$ by the constrained one given by:
\begin{equation}
r^{opt}(k)=\mbox{\rm arg}\min_{r} \hat J(r)\quad \mbox{\rm under $r_j=r_j^{operator}$} 
\end{equation} 
Nothing else has to be changed in the control architecture as explained in the previous sections. \\
\item In the second level, the operator decides to take the direct control of some actuator value, say the first control input $u_{11}$ of subsystem $S_1$ (see Figure \ref{handover}). Assume that before this decision the coupling signals between subsystems $S_1$ and $S_2$ where given by (see Figure \ref{handover}):
\begin{equation}
v_1=v_{12}\quad ;\quad v_2=v_{21}
\end{equation}   
Now when the operator decides to be directly involved, it comes obviously that the operator should be viewed as an additional new subsystem, say $S_3$ which acts on subsystem $S_1$ though the signal $u_{11}$\footnote{Another way to view the situation is to consider that the operator is embedded inside subsystem $S_2$ which now acts on subsystem $S_1$ from which the control input has been removed.}. Therefore, the operator handover at this level can be handled by two changes (see Figure \ref{handover}.(b)) :\\
\begin{enumerate}
\item First the subsystem $S_1$ looses the control input $u_{11}$,\\
\item $u_{11}$ is transformed into a new component of the coupling signal $v_1$, that is:
\begin{equation}
v_1= \begin{bmatrix}
v_{12}\cr u_{11}
\end{bmatrix}  \quad ;\quad v_2=v_{21}
\end{equation} 
\end{enumerate} 
\end{enumerate} 

Now this change sets obviously a new hierarchical framework that needs to be assessed and tuned using the same tools and concepts described above. In particular, it can be revealed that for some handover modes, the stability of the system might be threatened and/or the tuning of the filtering and the local weights need to be different. The contribution of this paper gives the theoretical background to rationally address this crucial industrial question. 
\section{Conclusion and further investigations} \label{secconclusion} 
In this paper, a complete framework is proposed for the design of a hierarchical control structure that addresses the issues of modularity, privacy and distributed computation that arise in the industrial process control context. The framework is presented in the case where MPC control design is used in the local level. However it goes without saying that the same framework can be used in the more common case where local PID controllers are used at the local level. Indeed, when PID controllers (or any other linear controllers) are used in the local level, the subsystems can still send the predictions of their coupling signals if the exogenous signals acting on them are given by the coordinator. This enables the convergence of the fixed-point iteration to be studied and the appropriate tuning changes to be undertaken if necessary. This enables the use of the proposed scheme in coordinating the overall consequences of the network of local controllers. 

Natural extension of the present work concerns constraints handling as well as experimental validation on the real process. Using the available nonlinear knowledge-based model is also under investigation. 
\bibliographystyle{plain}
\bibliography{biblio_hierarchical.bib}
\end{document}

%% file: spectrum.tex
%
%
%
\definecolor{mycolor1}{rgb}{0,0.447,0.741}%
\begin{tikzpicture}

\begin{axis}[%
width=0.5\textwidth,
height=0.2\textheight,
scale only axis,
separate axis lines,
every outer x axis line/.append style={darkgray!60!black},
every x tick label/.append style={font=\color{darkgray!60!black}},
xmin=0,
xmax=1,
xlabel={$\beta$},
xmajorgrids,
every outer y axis line/.append style={darkgray!60!black},
every y tick label/.append style={font=\color{darkgray!60!black}},
ymin=0.9,
ymax=1.2,
ymajorgrids,
title={Spectrum radius $\rho(Z(\beta))$ vs $\beta$}
]
\addplot [
color=mycolor1,
solid,
line width=2.0pt,
mark=o,
mark options={solid},
forget plot
]
table[row sep=crcr]{
0 1\\
0.05 0.995276348574839\\
0.1 0.990552697149675\\
0.15 0.985829045724518\\
0.2 0.981105394299352\\
0.25 0.976381742874197\\
0.3 0.971658091449038\\
0.35 0.966934440023874\\
0.4 0.962210788598709\\
0.45 0.95748713717355\\
0.5 0.952763485748398\\
0.55 0.965563382205777\\
0.6 0.997498871141382\\
0.65 1.03617962470699\\
0.7 1.08120258170368\\
0.75 1.13201737495384\\
0.8 1.18807276017395\\
0.85 1.24879435129521\\
0.9 1.31353827311868\\
0.95 1.38174350615052\\
1 1.45298584289384\\
};
\end{axis}
\end{tikzpicture}%